\documentclass[preprint,12pt]{elsarticle}

\usepackage{amssymb}
\usepackage{url,hyperref}

\biboptions{numbers,sort&compress}

\newcommand{\be}{\begin{equation}}
\newcommand{\ee}{\end{equation}}
\newcommand{\ba}{\begin{eqnarray}}
\newcommand{\ea}{\end{eqnarray}}

\newcounter{bla}

 \journal{Computer Physics Communications}

\begin{document}
 
 \begin{frontmatter}



\title{{\tt APFELgrid}: a high performance tool for parton density determinations}
\tnotetext[mytitlenote]{Preprints: OUTP-16-09P, CERN-TH-2016-103}


\author[a]{Valerio Bertone}
\author[b]{Stefano Carrazza}
\author[a]{Nathan P. Hartland\corref{author}}

\cortext[author] {Corresponding author.\\\textit{E-mail address:} nathan.hartland@physics.ox.ac.uk}
\address[a]{Rudolf Peierls Centre for Theoretical Physics,\\ 1 Keble Road, University of Oxford, OX1 3NP, Oxford, UK}
\address[b]{Theoretical Physics Department, CERN, Geneva, Switzerland}

\begin{abstract}
  We present a new software package designed to reduce the computational
  burden of hadron collider measurements in Parton Distribution Function (PDF)
  fits. The {\tt APFELgrid} package converts interpolated weight
  tables provided by {\tt APPLgrid} files into a more efficient format
  for PDF fitting by the combination with PDF and
  $\alpha_s$ evolution factors provided by {\tt APFEL}. This
  combination significantly reduces the number of operations required
  to perform the calculation of hadronic observables in PDF fits and
  simplifies the structure of the calculation into a readily optimised
  scalar product. We demonstrate that our technique can lead to a
  substantial speed improvement when compared to existing methods
  without any reduction in numerical accuracy.
\end{abstract}


\end{frontmatter}

 \newpage

\noindent {\Large \textbf{Program Summary}}\\
 
\begin{small}
\noindent
{\em Manuscript Title:} {\tt APFELgrid}: a high performance tool for parton density determinations                                      \\
{\em Authors:} V.~Bertone, S.~Carrazza, N.P.~Hartland                                               \\
{\em Program Title:} {\tt APFELgrid}                                          \\
{\em Journal Reference:}                                      \\
{\em Catalogue identifier:}                                   \\
{\em Licensing provisions:}   MIT license                                \\
{\em Programming language:}  C++                                 \\
{\em Computer:} PC/Mac                                               \\
{\em Operating system:} MacOS/Linux                                       \\
{\em RAM:} varying                                              
{\em Classification:}  11.1 General, High Energy Physics and Computing                                       \\
{\em External routines/libraries:}  {\tt APPLgrid}, {\tt APFEL}                          \\
{\em Nature of problem:}\\
 Fast computation of hadronic observables under the variation of parton distribution functions.
   \\
{\em Solution method:}\\
  Combination of interpolated weight grids from {\tt APPLgrid} files and evolution factors from {\tt APFEL} into efficient {\tt FastKernel} tables.\\
{\em Running time:} varying\\
   \\
\end{small}

\clearpage


\section{Introduction}\label{sec:intro}

Measurements at colliders such as the Tevatron and Large Hadron Collider (LHC) have a unique capacity to shed light upon the internal dynamics 
of the proton and provide constraints upon proton PDFs~\cite{Rojo:2015acz}. However including large hadron collider datasets 
in PDF fits can provide a significant challenge due to the large computational footprint of performing accurate
theoretical predictions over the many iterations required in a fitting procedure. In order to make the fullest use of current 
and future LHC results, efficient strategies for the computation of these observables must therefore be employed. 

The typical Monte Carlo software packages used to perform predictions for 
hadron collider observables cannot be easily deployed in a PDF fit due the processing time 
required to obtain accurate results (usually of the order of a few hours or more per data
point). To overcome such limitations, the typical strategy
adopted for fast cross section prediction relies on the precomputation
of the partonic hard cross sections in such a way that the standard numerical
convolution with any set of PDFs can be reliably approximated by means of
interpolation techniques.

Such interpolation strategies are implemented in the {\tt
  APPLgrid}~\cite{Carli:2010rw} and {\tt
  FastNLO}~\cite{Wobisch:2011ij} projects. For the computation of the
hard cross sections, these packages rely on external codes to which
they are interfaced by means of a suite of functions allowing for the
filling of PDF- and $\alpha_s$-independent look-up tables of
cross section weights. Monte Carlo programs such as {\tt MCFM}~\cite{Campbell:2010ff}
and {\tt NLOJet++}~\cite{Nagy:2003tz} have been interfaced directly to
{\tt APPLgrid}/{\tt FastNLO} and more recently de\-di\-ca\-ted
interfaces to automated general-purpose event
generators have been developed. The {\tt
  aMCfast}~\cite{Bertone:2014zva} and {\tt
  MCgrid}~\cite{DelDebbio:2013kxa} codes can generate interpolation grids in {\tt APPLgrid}/{\tt FastNLO} format 
 by extracting the relevant information from the {\tt
  MadGraph5\_aMC@NLO}~\cite{Alwall:2014hca} and {\tt
  SHERPA}~\cite{Gleisberg:2008ta} event generators respectively.

While these tools have proven to be invaluable in the extraction
of parton densities, the volume of experimental data made available by
LHC collaborations for use in PDF fits is already stretching the
capabilities of the typical fitting technology. A standard
global PDF fit may now include thousands of hadronic data points for which
predictions have to be computed thousands of times during the
minimisation process. As a consequence, performing these predictions
using the standard interpolating tools, $i.e.$
{\tt APPLgrid} and {\tt FastNLO}, starts to become prohibitively time-consuming.
For this reason a high-performance tool tailored specifically to the requirements of PDF analysis becomes increasingly
important.

The {\tt FastKernel} method was developed to
address this problem in the context of the NNPDF global analyses~\cite{Ball:2014uwa}. This method differs from the standard procedure
\`{a} la {\tt APPLgrid} or {\tt FastNLO} in that it maximises the
amount of information that is precomputed prior to fitting so as to
minimise the amount of operations required during the
fit. More specifically, the {\tt FastKernel} method relies on the
combination of precomputed hard cross sections with DGLAP evolution
kernels into a single look-up table, here called
a {\tt FastKernel} ({\tt FK}) table. In this way the prediction
for a given hadronic observable can be obtained by performing a simple matrix
product between the respective {\tt FK} table and PDFs evaluated directly at the fitting scale.

In this paper we present the {\tt APFELgrid} package, a public
implementation of the {\tt FastKernel} method in which the hard
partonic cross sections provided in an {\tt APPLgrid} look-up table
are combined with the DGLAP evolution kernels provided by the {\tt
  APFEL}
package~\cite{Bertone:2013vaa}.

This paper proceeds as follows. In
Sect.~\ref{sec:FastKernel} we present the technical details of the
implementation of the {\tt FastKernel} method. This is followed in
Sect.~\ref{sec:benchmark} by a performance benchmark of the
{\tt APFELgrid} library and resulting {\tt FK} tables. Finally, in
Sect.~\ref{sec:conclusion} we summarise the results discussed in this
work.

\section{Interpolation tools for collider observables}\label{sec:FastKernel}

Hadron collider observables are typically computed in QCD by means of a double
convolution of parton densities with a hard scattering
cross section. Consider for example the calculation of a general
cross section $pp\to X$ with a set of PDFs $\{f\}$:
\begin{equation}\label{eq:hadconv}
  \sigma_{pp\to X} =
  \sum_{s}\sum_{p} \int dx_1\,dx_2\,
  \hat{\sigma}^{(p)(s)}\,\alpha_s^{p+p_{\rm LO}}(Q^2) \,F^{(s)}(x_1,x_2, Q^2)\,,
\end{equation}
where $Q^2$ is the typical hard scale of the process, the index $s$
sums over the active partonic subprocesses in the calculation, $p$ sums over
the perturbative orders used in the expansion, $p_{\rm
  LO}$ is the leading-order power of $\alpha_s$ for the process and
$\hat{\sigma}^{(p)(s)}$ is the N$^p$LO contribution to the
cross section for the partonic subprocess scattering $(s)\to
X$. $F^{(s)}$ represents the subprocess parton density:
\begin{equation}\label{eq:APPLsubproc}
  F^{(s)}(x_1,x_2, Q^2) =\sum_{i,j} C^{(s)}_{ij} 
  f_i(x_{1},Q^2)f_j(x_{2},Q^2)\,,
\end{equation}
where the $C^{(s)}_{ij}$ matrix enumerates the combinations of PDFs
contributing to the $s$-th subprocess.  The central observation of
tools such as {\tt APPLgrid} and {\tt FastNLO} is that the PDF and
$\alpha_S$ dependence may be factorised out of the convolution via
expansion over a set of interpolating functions, spanning $Q^2$ and
the two values of parton-$x$. For example one may represent the
subprocess PDFs and $\alpha_S$ in terms of Lagrange basis
polynomials $\mathcal{I}_\tau(Q^2)$, $\mathcal{I}_\alpha(x_1)$ and
$\mathcal{I}_\beta(x_2)$ as:
\begin{equation}\label{eq:interpolation}
\begin{array}{c}
\displaystyle \alpha_s^{p+p_{\rm LO}}(Q^2) \,F^{(s)}(x_1,x_2, Q^2)
  =\\
\\
\displaystyle \sum_{\alpha,\beta,\tau} \alpha_s^{p+p_{\rm LO}}(Q_\tau^2)
  \,F^{(s)}_{\alpha\beta , \tau}
  \,\mathcal{I}_\tau(Q^2)\,\mathcal{I}_\alpha(x_1)
  \,\mathcal{I}_\beta(x_2),
\end{array}
\end{equation}
where we use the shorthand $F^{(s)}_{\alpha\beta ,\tau} =
F^{(s)}(x_\alpha, x_\beta,Q_\tau^2)$.  Using these expressions in the
double convolution of Eq.~(\ref{eq:hadconv}) one can finally obtain an
expression for the desired cross section which depends upon the
subprocess PDFs only through a simple product:

\begin{equation} \label{eq:applconv}
  \sigma_{pp\to X} = \sum_p \sum_{s}\sum_{\alpha,\beta,\tau} 
  \alpha_s^{p+p_{\rm LO}}(Q^2_\tau)W_{\alpha\beta,\tau}^{(p)(s)} \, F_{\alpha\beta,\tau}^{(s)}\,,
\end{equation}
where
\begin{equation} \label{eq:applgrid}
  W_{\alpha\beta,\tau}^{(p)(s)} = \mathcal{I}_\tau(Q^2)\int dx_1\,dx_2\,
  \hat{\sigma}^{(p)(s)}\,\mathcal{I}_\alpha(x_1)
  \,\mathcal{I}_\beta(x_2)\,,
\end{equation}
consists of the convolution of the hard cross section with the
interpolating polynomials. This information may be stored in a
precomputed look-up table.  The final expression for the cross section
in Eq.~(\ref{eq:applconv}) is therefore a considerably simpler task to perform
inside a fit than the direct evaluation of the double convolution.

\subsection{The FastKernel method}

A number of tools ($e.g.$ {\tt
  APFEL},
{\tt HOPPET}~\cite{Salam:2008qg} and {\tt QCDNUM}~\cite{Botje:2010ay}) 
are available which perform PDF evolution via an analogous
interpolation procedure. In such a way PDFs at a general scale $Q_\tau$ may be
expressed as a product of PDFs at some initial fitting scale $Q_0$ and an
\textit{evolution operator} obtained by the solution of the DGLAP equation.
\begin{equation}\label{eq:fastPDFfinal_recalled}
  f_i(x_{\alpha},Q^2_\tau) = \sum_{k}
  \sum_\beta A^\tau_{\alpha\beta, ik}\;
  f_k(x_\beta,Q^2_0)\,, 
\end{equation}
where latin indices run over PDF flavour, greek indices run over points in an initial-scale interpolating $x$-grid and
the evolution operator $A$ may be accessed directly in the {\tt APFEL} package. Given this operator, we may replace the
(general-scale) PDFs used in the subprocess parton density
Eq.~(\ref{eq:APPLsubproc}) with their equivalent expressions evaluated at the fitting scale as
\begin{equation}\label{eq:FK1}
\begin{array}{rcl}
F^{(s)}_{\alpha\beta,\tau} &=&  \displaystyle \sum_{i,j} \sum_{k,l}
                               \sum_{\delta,\gamma} C^{(s)}_{ij}
                               \left[  A^\tau_{\alpha\delta ik}\;
                               f_k(x_\delta,Q^2_0) A^\tau_{\beta\gamma
                               jl}\; f_l(x_\gamma,Q^2_0) \right]\;\;\;
  \\
\\
&=& \displaystyle \sum_{k,l}\sum_{\delta,\gamma}
\widetilde{C}^{(s),\tau}_{kl,\alpha\beta\gamma\delta}
f_k(x_\delta,Q^2_0) f_l(x_\gamma,Q^2_0)\,,
\end{array}
\end{equation}
with the object
\begin{equation}
  \widetilde{C}^{(s),\tau}_{kl,\alpha\beta\gamma\delta} =
  \sum_{i,j} C^{(s)}_{ij} A^\tau_{\alpha\delta ik}
  A^\tau_{\beta\gamma jl}\,,
\end{equation}
combining the operations of subprocess density construction and PDF
evolution. Going further and using the expression for
subprocess parton densities in Eq.~(\ref{eq:FK1}) in the full
cross section calculation we obtain
\begin{equation}
\sigma_{pp\to X} = \sum_{k,l}\sum_{\delta,\gamma}\sum_p
\sum_{s} \sum_{\alpha,\beta}
\sum_{\tau} 
\alpha_s^{p+p_{\rm LO}}(Q^2_\tau)W_{\alpha\beta,\tau}^{(p)(s)} \widetilde{C}^{(s),\tau}_{kl,\alpha\beta\gamma\delta}
f_k(x_\delta,Q^2_0) f_l(x_\gamma,Q^2_0)\,.
\end{equation}
Performing some further contractions it is possible to obtain an
extremely compact expression for the calculation of the cross section
in question, in terms of only the initial-scale PDFs and summing only
over the initial scale interpolating $x$-grid and the incoming parton
flavours:
\begin{equation}\label{eq:FK}
  \sigma _{pp\to X} = \sum_{k,l}\sum_{\delta,\gamma} 
  \widetilde{W}_{kl,\delta\gamma} \,f_k(x_\delta,Q^2_0) f_l(x_\gamma,Q^2_0),
\end{equation}
where the object:
\begin{equation}\label{eq:FKTable}
  \widetilde{W}_{kl,\delta\gamma} = \sum_p\sum_{s}\sum_{\alpha,\beta} \sum_{\tau}
\alpha_s^{p+p_{\rm LO}}(Q^2_\tau)  W_{\alpha\beta,\tau}^{(p)(s)} \widetilde{C}^{(s),\tau}_{kl,\alpha\beta\gamma\delta}
\end{equation}
is referred to here as an {\tt FK} table, and combines the information
stored in {\tt APPLgrid}-style interpolated weight grids with
analogously interpolated DGLAP evolution operators. This combination
enables for a maximally efficient expression for the
calculation of observables at hadron colliders under PDF variation, without invoking
any additional approximation. 

\subsection{Features and limitations of {\tt FK} tables}

The {\tt FK} product of Eq.~(\ref{eq:FK}) differs with respect to the
product in Eq.~(\ref{eq:applconv}) in several notable ways. Firstly
the typical {\tt APPLgrid} or {\tt FastNLO} products use as input PDFs at a
general scale, requiring that PDF evolution {\it e.g.}
Eq.~(\ref{eq:fastPDFfinal_recalled}) be performed for every variation of the PDFs
during the fit. In the {\tt FK} product
this evolution is pre-cached at the stage of {\tt FK} table
generation, requiring only initial-scale PDFs at the time of fitting. 
This pre-caching of the evolution also removes the need to
 sum over hard scale and perturbative order during the
fit, further reducing the number of operations required. As the {\tt FK}
product acts directly at the fitting scale, it benefits from the typically reduced number of active
partonic modes, with the sum over flavours in
Eq.~(\ref{eq:FK}) being limited to those directly parametrised in the
fit. Having reduced the calculation to such a simple form, it is also
straightforward to apply standard 
computational tools such as multi-threading through e.g OpenMP or
Single Instruction Multiple Data (SIMD) operations such as SSE or AVX to
further reduce computational expense. 

While these features provide significant performance enhancements, the
{\tt FK} table format is not suitable as a complete replacement for
tools such as {\tt APPLgrid}. The precomputation of the PDF evolution
necessarily means that all theory parameters such as perturbative
order, strong coupling and factorization/renormalization scales are
inextricably embedded in each {\tt FK} table. In order to perform PDF
fits including variations of these parameters, multiple {\tt FK}
tables must be computed, each with different theory settings. While
performing such a re-calculation directly from Monte Carlo codes would
be exceptionally time consuming, the data representation in {\tt
  APPLgrid} files allows for an efficient (re)-combination with varying
theory parameters.

\section{Performance benchmarks}
\label{sec:benchmark}

We shall now examine the performance differences between
the two expressions for fast interpolated cross section prediction
Eq.~(\ref{eq:applconv}) ({\tt APPLgrid}) and Eq.~(\ref{eq:FK}) ({\tt
  FK}).
In order to provide a comprehensive benchmark, we consider here a wide range of
processes including LHC and Tevatron electroweak vector boson
production
measurements~\cite{Aaij:2012mda,Aaij:2012vn,Chatrchyan:2013mza,Chatrchyan:2013uja,Chatrchyan:2012xt,Aad:2013iua,Aad:2011fp,Aad:2011dm,Aaltonen:2010zza},
$t\bar{t}$ total
cross sections~\cite{ATLAS:2012aa, Chatrchyan:2013faa,Chatrchyan:2012bra,Chatrchyan:2012ria},
double-differential Drell-Yan
cross sections~\cite{Chatrchyan:2013tia,CMS:2014jea} and inclusive jet
data~\cite{Chatrchyan:2012bja,Aad:2011fc,Aad:2013lpa,Abazov:2007jy}. Predictions
are performed over a wide range of kinematics, for a total of 52
source {\tt APPLgrid} files corresponding to the majority of available
LHC and Tevatron datasets applicable to PDF determination.
While the source {\tt APPLgrid} files have varying grid densities in
$x$ and $Q^2$, for the purposes of comparison the corresponding {\tt
  FK} tables are produced consistently with 30 points in $x$, and at
an initial scale below the charm threshold, therefore with seven
active partonic species. These settings are chosen so as to provide a realistic
comparison, in a production environment the density and distribution of the
$x$-grid may be adjusted to match interpolation accuracy requirements.  For these comparisons the {\tt FK}
table is stored as double-precision in memory for table generation and in
single-precision for the purposes of computing the {\tt FK} product.

\begin{figure}
  \centering
  \includegraphics[scale=0.6]{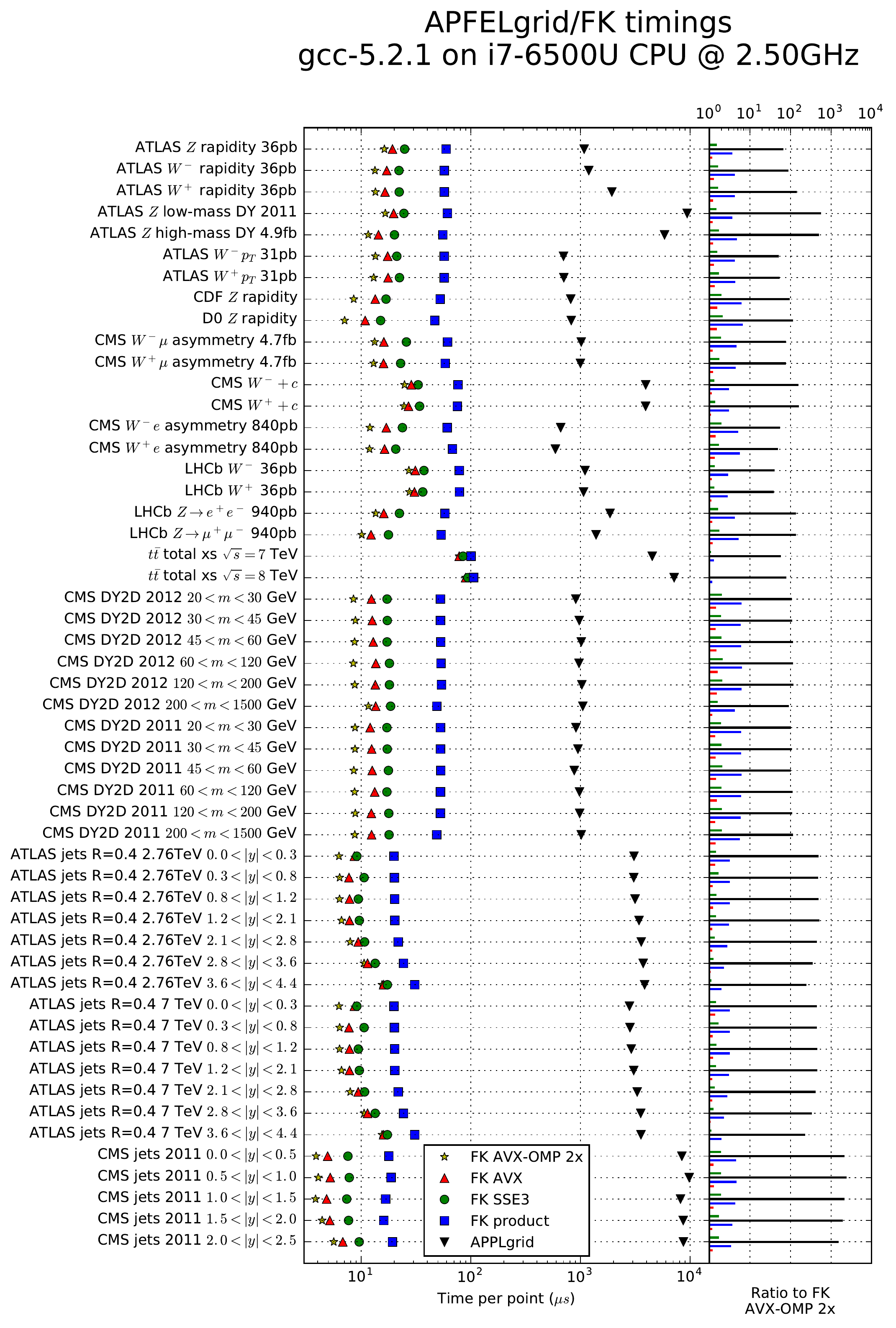}
\caption{\small Performance comparisons between {\tt FK} with
  AVX-OpenMP 2x, AVX, SSE3, double precision convolution and {\tt
    APPLgrid} convolution time per point and process.}
\label{fig:timings}
\end{figure}

In Fig.~\ref{fig:timings} we compare the average time taken per
datapoint for the {\tt FK} and {\tt APPLgrid} calculations, for all of
the 52 tables. We show timings for the {\tt FK} calculation in four
different configurations: AVX-OpenMP 2x (2 CPU cores), AVX, SSE3 and
the standard double precision product. Due to the inherent structural
differences between the {\tt FK} and {\tt APPLgrid} procedures,
results from the {\tt FK} calculation are systematically faster than
those from {\tt APPLgrid}. In particular, when comparing {\tt
  FK} AVX-OpenMP 2x to {\tt APPLgrid} timings we obtain minimally a factor of
ten improvement in speed for electroweak vector boson production and a
maximum factor of 2000 improvement in predictions for inclusive jet
data. Across all processes and kinematic regions we observe
significant performance improvements from using the {\tt FK}
calculation even without the use of SIMD or multi-threading.

While sheer computational speed is typically the primary consideration
when com\-pu\-ting observables in a PDF fit, other factors such as
table size in the filesystem and memory, along with the computational
cost of pre-computing {\tt FK} tables must be considered. Indeed, the
computation of the {\tt FK} table in Eq.~(\ref{eq:FKTable}) requires a
great deal of operations which can be time consuming, particularly in
the case of source {\tt APPLgrid} files with very high interpolation
precision.

\begin{figure}
  \centering                                      
  \includegraphics[scale=0.6]{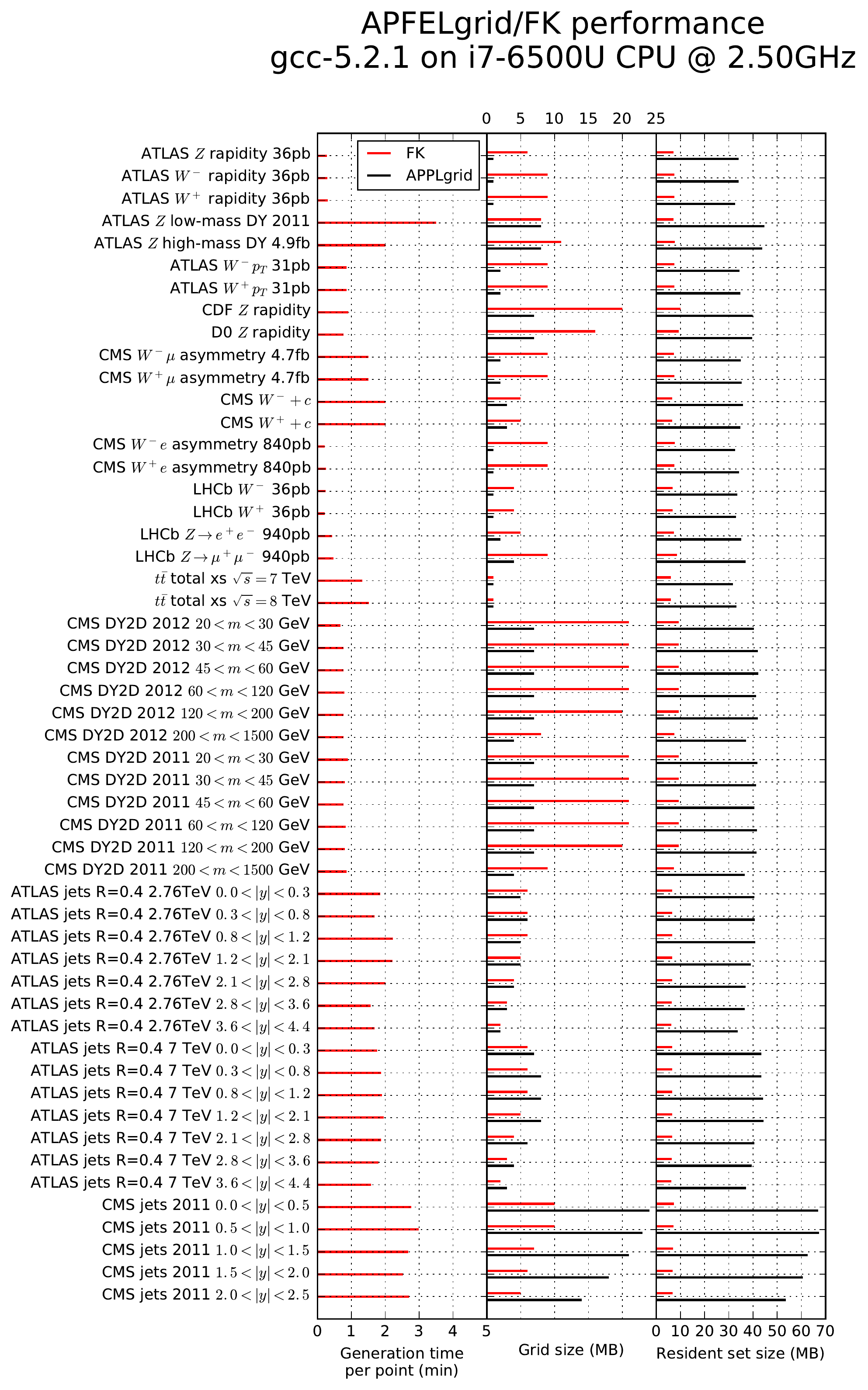}
\caption{\small {\tt APFELgrid} generation time per point and
  comparison to {\tt APPLgrid} for grid size on disk and resident set size
  (RSS).}
\label{fig:performance}
\end{figure}

In Fig.~\ref{fig:performance} we examine the {\tt FK} table generation
time with {\tt APFELgrid}, {\tt FK} table file size and memory usage
of the {\tt FK} tables arising from the same source {\tt APPLgrid}
files as discussed in Fig.~\ref{fig:timings}. When examining the
table generation time per point, we observe timings from a few
milliseconds to 3.5 minutes per point, with differences arising from
the varying grid densities used in the input {\tt APPLgrid} files.  In
terms of the grid size on disk, {\tt FK} tables are typically larger
than their corresponding {\tt APPLgrid} files, primarily as the {\tt
  FK} file format is encoded in plain text for compatibility whereas
{\tt APPLgrid} files are expressed in binary as {\tt ROOT}
files. However when measuring the in-memory resident set size used by
the two procedures, the amount of system memory used by {\tt
  FK} tables is systematically less than {\tt APPLgrid} files for all
processes considered here by at least 75\%. Note that this effect is
in part due to the differing precisions of the default representations.


\section{Conclusion}\label{sec:conclusion}

In this work we have demonstrated that by employing the so-called {\tt
  FastKernel} method, it is possible to convert an {\tt APPLgrid}
weight table into a derived format, referred to as an {\tt FK} table,
including the effects of PDF and $\alpha_s$ evolution. This procedure has been
implemented in the {\tt APFELgrid} package, supplied as a set of {\tt
  C++} routines designed to supplement the PDF evolution library {\tt
  APFEL} with {\tt FK} table generation capabilities.
The {\tt APFELgrid} package allows one to obtain a computationally
efficient expression for the calculation of hadronic cross sections,
in terms of only the initial-scale PDFs and summing only over the
initial scale interpolating $x$-grid and the incoming parton
flavours. The simple structure of the resulting
product makes {\tt FK} tables particularly suitable for the
efficient use of computational tools such as {\tt SIMD} and {\tt
  OpenMP}.

We have shown that in several practical examples the numerical
evaluation of an {\tt FK} product is considerably faster than the
corresponding {\tt APPLgrid} product, even in the case where neither
SIMD or multi-threading are applied. {\tt FK} tables are supplied in a
simple plain-text format in order to simplify the construction of user
interfaces, and therefore are typically larger than 
corresponding {\tt APPLgrids}. However we have shown that the
in-memory resident set sizes occupied by {\tt FK} tables are typically
smaller than those required by {\tt APPLgrids}, in our examples by at least 75\%.

The substantial speed improvement of {\tt FK} tables with respect to
{\tt APPLgrid} in association with the reduction in memory footprint
makes the {\tt APFELgrid} code a valuable tool for modern PDF fits
including large collider datasets.

The {\tt APFELgrid} package and associated documentation are publicly
available from the webpage:
\begin{center}
\url{https://github.com/nhartland/APFELgrid}
\end{center}

\section*{Acknowledgements}

The authors would like to thank members of the NNPDF Collaboration for their support and
motivation for this work; particularly Juan Rojo, Luigi del Debbio and Alberto Guffanti. We would also like
to thank Mark Sutton for helpful comments on the paper. V.~B. and N.~H. are
supported by an European Research Council Starting Grant ``PDF4BSM''.
S.~C. is supported by the HICCUP ERC Consolidator grant (614577).




\bibliographystyle{elsarticle-num}
\bibliography{paper}

\begin{thebibliography}{10}
\expandafter\ifx\csname url\endcsname\relax
  \def\url#1{\texttt{#1}}\fi
\expandafter\ifx\csname urlprefix\endcsname\relax\def\urlprefix{URL }\fi
\expandafter\ifx\csname href\endcsname\relax
  \def\href#1#2{#2} \def\path#1{#1}\fi

\bibitem{Rojo:2015acz}
J.~Rojo, et~al., {The PDF4LHC report on PDFs and LHC data: Results from Run I
  and preparation for Run II}, J. Phys. G42 (2015) 103103.
\newblock \href {http://arxiv.org/abs/1507.00556} {\path{arXiv:1507.00556}},
  \href {http://dx.doi.org/10.1088/0954-3899/42/10/103103}
  {\path{doi:10.1088/0954-3899/42/10/103103}}.

\bibitem{Carli:2010rw}
T.~Carli, et~al., {A posteriori inclusion of parton density functions in NLO
  QCD final-state calculations at hadron colliders: The APPLGRID Project},
  Eur.Phys.J. C66 (2010) 503.
\newblock \href {http://arxiv.org/abs/0911.2985} {\path{arXiv:0911.2985}},
  \href {http://dx.doi.org/10.1140/epjc/s10052-010-1255-0}
  {\path{doi:10.1140/epjc/s10052-010-1255-0}}.

\bibitem{Wobisch:2011ij}
M.~Wobisch, D.~Britzger, T.~Kluge, K.~Rabbertz, F.~Stober, {Theory-Data
  Comparisons for Jet Measurements in Hadron-Induced Processes}\href
  {http://arxiv.org/abs/1109.1310} {\path{arXiv:1109.1310}}.

\bibitem{Campbell:2010ff}
J.~M. Campbell, R.~K. Ellis, {MCFM for the Tevatron and the LHC}, Nucl. Phys.
  Proc. Suppl. 205-206 (2010) 10--15.
\newblock \href {http://arxiv.org/abs/1007.3492} {\path{arXiv:1007.3492}},
  \href {http://dx.doi.org/10.1016/j.nuclphysbps.2010.08.011}
  {\path{doi:10.1016/j.nuclphysbps.2010.08.011}}.

\bibitem{Nagy:2003tz}
Z.~Nagy, {Next-to-leading order calculation of three jet observables in hadron
  hadron collision}, Phys. Rev. D68 (2003) 094002.
\newblock \href {http://arxiv.org/abs/hep-ph/0307268}
  {\path{arXiv:hep-ph/0307268}}, \href
  {http://dx.doi.org/10.1103/PhysRevD.68.094002}
  {\path{doi:10.1103/PhysRevD.68.094002}}.

\bibitem{Bertone:2014zva}
V.~Bertone, R.~Frederix, S.~Frixione, J.~Rojo, M.~Sutton, {aMCfast: automation
  of fast NLO computations for PDF fits}, JHEP 08 (2014) 166.
\newblock \href {http://arxiv.org/abs/1406.7693} {\path{arXiv:1406.7693}},
  \href {http://dx.doi.org/10.1007/JHEP08(2014)166}
  {\path{doi:10.1007/JHEP08(2014)166}}.

\bibitem{DelDebbio:2013kxa}
L.~Del~Debbio, N.~P. Hartland, S.~Schumann, {MCgrid: projecting cross section
  calculations on grids}, Comput. Phys. Commun. 185 (2014) 2115--2126.
\newblock \href {http://arxiv.org/abs/1312.4460} {\path{arXiv:1312.4460}},
  \href {http://dx.doi.org/10.1016/j.cpc.2014.03.023}
  {\path{doi:10.1016/j.cpc.2014.03.023}}.

\bibitem{Alwall:2014hca}
J.~Alwall, R.~Frederix, S.~Frixione, V.~Hirschi, F.~Maltoni, O.~Mattelaer,
  H.~S. Shao, T.~Stelzer, P.~Torrielli, M.~Zaro, {The automated computation of
  tree-level and next-to-leading order differential cross sections, and their
  matching to parton shower simulations}, JHEP 07 (2014) 079.
\newblock \href {http://arxiv.org/abs/1405.0301} {\path{arXiv:1405.0301}},
  \href {http://dx.doi.org/10.1007/JHEP07(2014)079}
  {\path{doi:10.1007/JHEP07(2014)079}}.

\bibitem{Gleisberg:2008ta}
T.~Gleisberg, S.~Hoeche, F.~Krauss, M.~Schonherr, S.~Schumann, F.~Siegert,
  J.~Winter, {Event generation with SHERPA 1.1}, JHEP 02 (2009) 007.
\newblock \href {http://arxiv.org/abs/0811.4622} {\path{arXiv:0811.4622}},
  \href {http://dx.doi.org/10.1088/1126-6708/2009/02/007}
  {\path{doi:10.1088/1126-6708/2009/02/007}}.

\bibitem{Ball:2014uwa}
R.~D. Ball, et~al., {Parton distributions for the LHC Run II}, JHEP 04 (2015)
  040.
\newblock \href {http://arxiv.org/abs/1410.8849} {\path{arXiv:1410.8849}},
  \href {http://dx.doi.org/10.1007/JHEP04(2015)040}
  {\path{doi:10.1007/JHEP04(2015)040}}.

\bibitem{Bertone:2013vaa}
V.~Bertone, S.~Carrazza, J.~Rojo, {APFEL: A PDF Evolution Library with QED
  corrections}, Comput. Phys. Commun. 185 (2014) 1647--1668.
\newblock \href {http://arxiv.org/abs/1310.1394} {\path{arXiv:1310.1394}},
  \href {http://dx.doi.org/10.1016/j.cpc.2014.03.007}
  {\path{doi:10.1016/j.cpc.2014.03.007}}.

\bibitem{Salam:2008qg}
G.~P. Salam, J.~Rojo, {A Higher Order Perturbative Parton Evolution Toolkit
  (HOPPET)}, Comput. Phys. Commun. 180 (2009) 120--156.
\newblock \href {http://arxiv.org/abs/0804.3755} {\path{arXiv:0804.3755}},
  \href {http://dx.doi.org/10.1016/j.cpc.2008.08.010}
  {\path{doi:10.1016/j.cpc.2008.08.010}}.

\bibitem{Botje:2010ay}
M.~Botje, {QCDNUM: Fast QCD Evolution and Convolution}, Comput. Phys. Commun.
  182 (2011) 490--532.
\newblock \href {http://arxiv.org/abs/1005.1481} {\path{arXiv:1005.1481}},
  \href {http://dx.doi.org/10.1016/j.cpc.2010.10.020}
  {\path{doi:10.1016/j.cpc.2010.10.020}}.

\bibitem{Aaij:2012mda}
R.~Aaij, et~al., {Measurement of the cross-section for $Z \to e^+e^-$
  production in $pp$ collisions at $\sqrt{s}=7$ TeV}, JHEP 02 (2013) 106.
\newblock \href {http://arxiv.org/abs/1212.4620} {\path{arXiv:1212.4620}},
  \href {http://dx.doi.org/10.1007/JHEP02(2013)106}
  {\path{doi:10.1007/JHEP02(2013)106}}.

\bibitem{Aaij:2012vn}
R.~Aaij, et~al., {Inclusive $W$ and $Z$ production in the forward region at
  $\sqrt{s} = 7$ TeV}, JHEP 06 (2012) 058.
\newblock \href {http://arxiv.org/abs/1204.1620} {\path{arXiv:1204.1620}},
  \href {http://dx.doi.org/10.1007/JHEP06(2012)058}
  {\path{doi:10.1007/JHEP06(2012)058}}.

\bibitem{Chatrchyan:2013mza}
S.~Chatrchyan, et~al., {Measurement of the muon charge asymmetry in inclusive
  $pp \to W+X$ production at $\sqrt s =$ 7 TeV and an improved determination of
  light parton distribution functions}, Phys. Rev. D90~(3) (2014) 032004.
\newblock \href {http://arxiv.org/abs/1312.6283} {\path{arXiv:1312.6283}},
  \href {http://dx.doi.org/10.1103/PhysRevD.90.032004}
  {\path{doi:10.1103/PhysRevD.90.032004}}.

\bibitem{Chatrchyan:2013uja}
S.~Chatrchyan, et~al., {Measurement of associated W + charm production in pp
  collisions at $\sqrt{s}$ = 7 TeV}, JHEP 02 (2014) 013.
\newblock \href {http://arxiv.org/abs/1310.1138} {\path{arXiv:1310.1138}},
  \href {http://dx.doi.org/10.1007/JHEP02(2014)013}
  {\path{doi:10.1007/JHEP02(2014)013}}.

\bibitem{Chatrchyan:2012xt}
S.~Chatrchyan, et~al., {Measurement of the electron charge asymmetry in
  inclusive $W$ production in $pp$ collisions at $\sqrt{s}=7$ TeV}, Phys. Rev.
  Lett. 109 (2012) 111806.
\newblock \href {http://arxiv.org/abs/1206.2598} {\path{arXiv:1206.2598}},
  \href {http://dx.doi.org/10.1103/PhysRevLett.109.111806}
  {\path{doi:10.1103/PhysRevLett.109.111806}}.

\bibitem{Aad:2013iua}
G.~Aad, et~al., {Measurement of the high-mass Drell--Yan differential
  cross-section in pp collisions at $\sqrt{s}=7$ TeV with the ATLAS detector},
  Phys. Lett. B725 (2013) 223--242.
\newblock \href {http://arxiv.org/abs/1305.4192} {\path{arXiv:1305.4192}},
  \href {http://dx.doi.org/10.1016/j.physletb.2013.07.049}
  {\path{doi:10.1016/j.physletb.2013.07.049}}.

\bibitem{Aad:2011fp}
G.~Aad, et~al., {Measurement of the Transverse Momentum Distribution of $W$
  Bosons in $pp$ Collisions at $\sqrt{s}=7$ TeV with the ATLAS Detector}, Phys.
  Rev. D85 (2012) 012005.
\newblock \href {http://arxiv.org/abs/1108.6308} {\path{arXiv:1108.6308}},
  \href {http://dx.doi.org/10.1103/PhysRevD.85.012005}
  {\path{doi:10.1103/PhysRevD.85.012005}}.

\bibitem{Aad:2011dm}
G.~Aad, et~al., {Measurement of the inclusive $W^\pm$ and $Z/\gamma$ cross
  sections in the electron and muon decay channels in $pp$ collisions at
  $\sqrt{s}=7$ TeV with the ATLAS detector}, Phys. Rev. D85 (2012) 072004.
\newblock \href {http://arxiv.org/abs/1109.5141} {\path{arXiv:1109.5141}},
  \href {http://dx.doi.org/10.1103/PhysRevD.85.072004}
  {\path{doi:10.1103/PhysRevD.85.072004}}.

\bibitem{Aaltonen:2010zza}
T.~A. Aaltonen, et~al., {Measurement of $d\sigma/dy$ of Drell-Yan $e^+e^-$
  pairs in the $Z$ Mass Region from $p\bar{p}$ Collisions at $\sqrt{s}=1.96$
  TeV}, Phys. Lett. B692 (2010) 232--239.
\newblock \href {http://arxiv.org/abs/0908.3914} {\path{arXiv:0908.3914}},
  \href {http://dx.doi.org/10.1016/j.physletb.2010.06.043}
  {\path{doi:10.1016/j.physletb.2010.06.043}}.

\bibitem{ATLAS:2012aa}
G.~Aad, et~al., {Measurement of the cross section for top-quark pair production
  in $pp$ collisions at $\sqrt{s}=7$ TeV with the ATLAS detector using final
  states with two high-pt leptons}, JHEP 05 (2012) 059.
\newblock \href {http://arxiv.org/abs/1202.4892} {\path{arXiv:1202.4892}},
  \href {http://dx.doi.org/10.1007/JHEP05(2012)059}
  {\path{doi:10.1007/JHEP05(2012)059}}.

\bibitem{Chatrchyan:2013faa}
S.~Chatrchyan, et~al., {Measurement of the $t \bar{t}$ production cross section
  in the dilepton channel in pp collisions at $\sqrt{s}$ = 8 TeV}, JHEP 02
  (2014) 024, [Erratum: JHEP02,102(2014)].
\newblock \href {http://arxiv.org/abs/1312.7582} {\path{arXiv:1312.7582}},
  \href {http://dx.doi.org/10.1007/JHEP02(2014)024, 10.1007/JHEP02(2014)102}
  {\path{doi:10.1007/JHEP02(2014)024, 10.1007/JHEP02(2014)102}}.

\bibitem{Chatrchyan:2012bra}
S.~Chatrchyan, et~al., {Measurement of the $t\bar{t}$ production cross section
  in the dilepton channel in $pp$ collisions at $\sqrt{s}=7$ TeV}, JHEP 11
  (2012) 067.
\newblock \href {http://arxiv.org/abs/1208.2671} {\path{arXiv:1208.2671}},
  \href {http://dx.doi.org/10.1007/JHEP11(2012)067}
  {\path{doi:10.1007/JHEP11(2012)067}}.

\bibitem{Chatrchyan:2012ria}
S.~Chatrchyan, et~al., {Measurement of the $t\bar{t}$ production cross section
  in $pp$ collisions at $\sqrt{s}=7$ TeV with lepton + jets final states},
  Phys. Lett. B720 (2013) 83--104.
\newblock \href {http://arxiv.org/abs/1212.6682} {\path{arXiv:1212.6682}},
  \href {http://dx.doi.org/10.1016/j.physletb.2013.02.021}
  {\path{doi:10.1016/j.physletb.2013.02.021}}.

\bibitem{Chatrchyan:2013tia}
S.~Chatrchyan, et~al., {Measurement of the differential and double-differential
  Drell-Yan cross sections in proton-proton collisions at $\sqrt{s} =$ 7 TeV},
  JHEP 12 (2013) 030.
\newblock \href {http://arxiv.org/abs/1310.7291} {\path{arXiv:1310.7291}},
  \href {http://dx.doi.org/10.1007/JHEP12(2013)030}
  {\path{doi:10.1007/JHEP12(2013)030}}.

\bibitem{CMS:2014jea}
V.~Khachatryan, et~al., {Measurements of differential and double-differential
  Drell-Yan cross sections in proton-proton collisions at 8 TeV}, Eur. Phys. J.
  C75~(4) (2015) 147.
\newblock \href {http://arxiv.org/abs/1412.1115} {\path{arXiv:1412.1115}},
  \href {http://dx.doi.org/10.1140/epjc/s10052-015-3364-2}
  {\path{doi:10.1140/epjc/s10052-015-3364-2}}.

\bibitem{Chatrchyan:2012bja}
S.~Chatrchyan, et~al., {Measurements of differential jet cross sections in
  proton-proton collisions at $\sqrt{s}=7$ TeV with the CMS detector}, Phys.
  Rev. D87~(11) (2013) 112002, [Erratum: Phys. Rev.D87,no.11,119902(2013)].
\newblock \href {http://arxiv.org/abs/1212.6660} {\path{arXiv:1212.6660}},
  \href {http://dx.doi.org/10.1103/PhysRevD.87.112002,
  10.1103/PhysRevD.87.119902} {\path{doi:10.1103/PhysRevD.87.112002,
  10.1103/PhysRevD.87.119902}}.

\bibitem{Aad:2011fc}
G.~Aad, et~al., {Measurement of inclusive jet and dijet production in $pp$
  collisions at $\sqrt{s}=7$ TeV using the ATLAS detector}, Phys. Rev. D86
  (2012) 014022.
\newblock \href {http://arxiv.org/abs/1112.6297} {\path{arXiv:1112.6297}},
  \href {http://dx.doi.org/10.1103/PhysRevD.86.014022}
  {\path{doi:10.1103/PhysRevD.86.014022}}.

\bibitem{Aad:2013lpa}
G.~Aad, et~al., {Measurement of the inclusive jet cross section in pp
  collisions at $\sqrt{s}=2.76$ TeV and comparison to the inclusive jet cross
  section at $\sqrt{s}=7$ TeV using the ATLAS detector}, Eur. Phys. J. C73~(8)
  (2013) 2509.
\newblock \href {http://arxiv.org/abs/1304.4739} {\path{arXiv:1304.4739}},
  \href {http://dx.doi.org/10.1140/epjc/s10052-013-2509-4}
  {\path{doi:10.1140/epjc/s10052-013-2509-4}}.

\bibitem{Abazov:2007jy}
V.~M. Abazov, et~al., {Measurement of the shape of the boson rapidity
  distribution for $p \bar{p} \to Z/\gamma^* \to e^{+} e^{-}$ + $X$ events
  produced at $\sqrt{s}$ of 1.96-TeV}, Phys. Rev. D76 (2007) 012003.
\newblock \href {http://arxiv.org/abs/hep-ex/0702025}
  {\path{arXiv:hep-ex/0702025}}, \href
  {http://dx.doi.org/10.1103/PhysRevD.76.012003}
  {\path{doi:10.1103/PhysRevD.76.012003}}.

\end{thebibliography}







\end{document}